\documentclass{article}
\usepackage{enumitem}
\setlist{nosep}

\usepackage{multicol}
\usepackage{graphicx}
\usepackage{sidecap}
\usepackage{amsmath}
\usepackage{siunitx}
\usepackage{wrapfig}
\usepackage{pifont}

\usepackage{tikz}

\newcommand*\numcircledmod[1]{\raisebox{.5pt}{\textcircled{\raisebox{-.9pt} {#1}}}}

\title{A simplified superheating Rankine pump with possible application in irrigation}

\author{M.-F. Danca}

\begin{document}
\maketitle
\begin{abstract}
In this paper a simplified superheating Rankine pump is presented. For an easy understanding and study of its functioning principle, the pump has been experimentally implemented, with superheating obtained by an electrical resistance. The pump can be adapted to serve as a reliable and cheap irrigation pump, having reduced maintenance and operating costs, and consuming clean energy (concentrate solar energy).
\end{abstract}
\vspace{3mm}
\textbf{keywords} Superheating, Rankine cycle, irrigation system
\section{Introduction}

As is well known, superheating appears when a liquid is heated under a temperature above its boiling point without boiling (vaporization) (see e.g. \cite{axa,unu,trei,patru,cinci,sase,sapte,opt,zece}). The superheating can determine a dramatic increase of vapor volume. For example, by super heating one gram of pure water from $100^\circ C$  to $100.26^\circ C$, one obtains $1.3$ l of vapor.
Generally speaking, a substance undergoes a phase change from the liquid state to the gaseous state while it is heated to its boiling point. As is well known, pure water boils at $100^\circ C$ under standard atmospheric pressure. However, if one considers a container with a smooth surface, such as a glass, the relatively static heating environment inside a microwave oven is unfavorable for the formation of steam bubbles. Even it is heated to or above its boiling point, the water is prevented from converting into steam and thus the boiling process is delayed. The water is said to be in a superheated state.

On the other side, there are several hydraulic pumps that use solar energy. Generally, these solutions are based on a solar panel that generates electricity and a control system used to drive a submersible water pump. The disadvantage is the high cost of installation due to system complexity and low efficiency due to successive conversion of solar energy into electricity-mechanical-hydraulic. Another more complicated system, using solar energy, consists of a concentrator solar radiation that radiates some horizontal water pipes, resulting saturated vapor passing through a solar superheater which is formed by another set of horizontal pipes \cite{12}. Transforming solar energy into steam can be achieved by a system composed of a central receiver and a superheater. Solar radiation is concentrated by a heliostat radiation on exposed surfaces of the evaporator and superheater, resulting in superheated steam \cite{13}. These systems have certain disadvantages in the required irrigation water pumping operations. Also, most irrigation systems are gravitational \cite{14,15}.

In this paper, a simplified traditional superheating Rankine-like pump is proposed (see e.g. \cite{unu,opt,17} and references therein). The pump uses water as the working fluid, which might transport water at higher destination than the level of the water source. Compared to the existing heating pumps, which usually have closed water and vapor circuits, the exhaust end of the proposed pump is open and, therefore, the system works under the exterior pressure. However, as shown in Section 3, it can be considered as a closed-loop system, where a working fluid repeatedly circulates through its components.

The paper is organized as follows: Section 2 describes the pump concept, Section 3 presents the underlying related phenomena, Section 4 proposes the possible adaptation for irrigation, and Conclusion section ends the investigation.

\section{The pump}
The proposed system is a simplified superheated Rankine-like cycle, the fundamental operating cycle, where an operating fluid is continuously evaporated and condensed (see e.g., \cite{opt}), with a ``liquid piston''.

\begin{figure}
\includegraphics[scale=0.33]{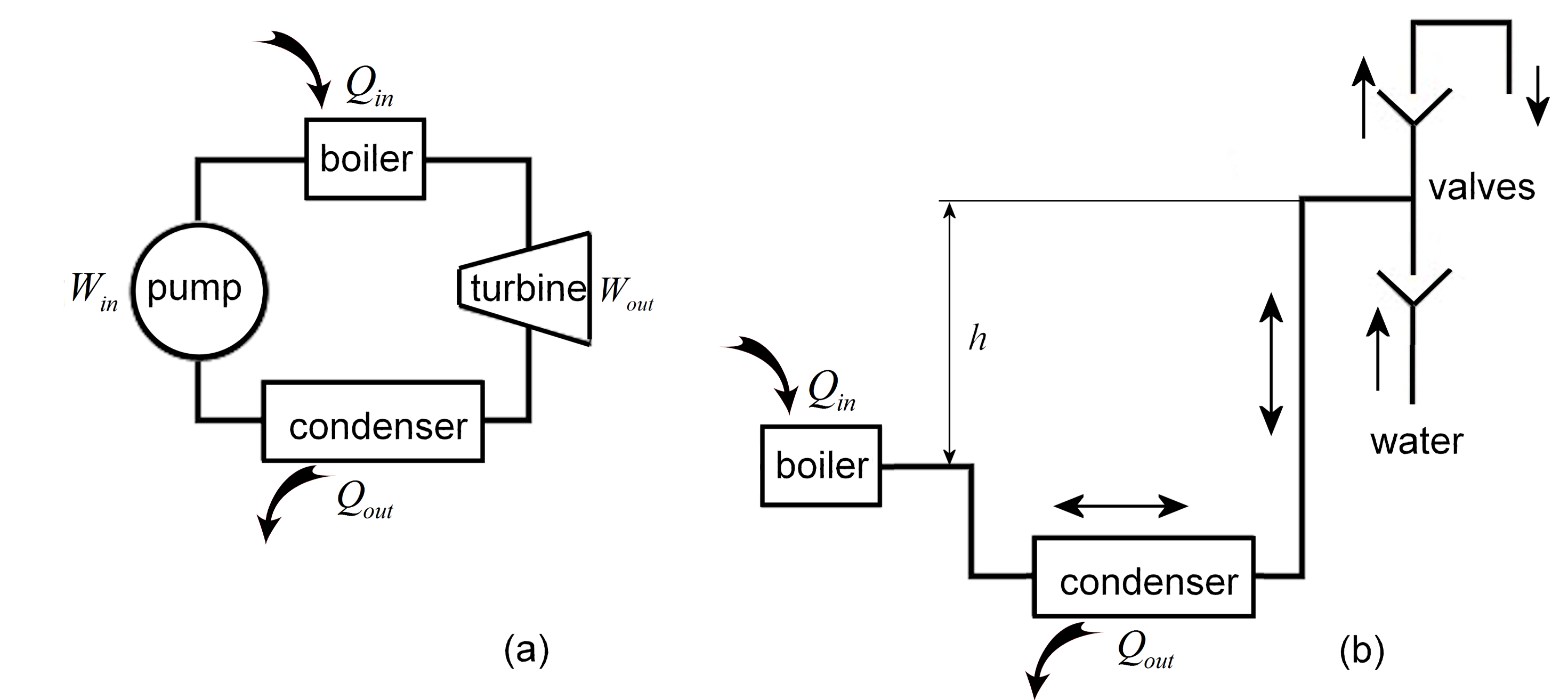}
\caption{(a) Sketch of a standard Rankine cycle; (b) Sketch of the proposed simplified Rankine pump.}
\label{fig0}
\end{figure}

A classical Rankine superheating cycle, where the working fluid undergos the phase change from a liquid to vapor phase, and vice versa, is composed of four main devices: pump, boiler, turbine and condenser (Fig. \ref{fig0} (a)). Superheating is important because it increases the thermal efficiency of the Rankine cycle.

In this paper, for simplicity, by \emph{pump} one understands the proposed simplified superheating Rankine cycle system.

The system is composed of only three devices: evaporator (boiler), condenser and a valves subsystem situated at a hight $h$ from evaporator (Fig. \ref{fig0} (b)). The role of the pump in the classical Rankine system is replaced here by the influence of gravitational force applied to the column of water with weight $h$ (gravitational water). Actually, one can consider that the proposed system still has a pump-like (the water column in the exhaust will pump, via the liquid piston, getting the water back in evaporator). Also, there exists a turbine-like (water transfer within the valves subsystem) structure.
\begin{SCfigure}
\includegraphics[scale=0.3]{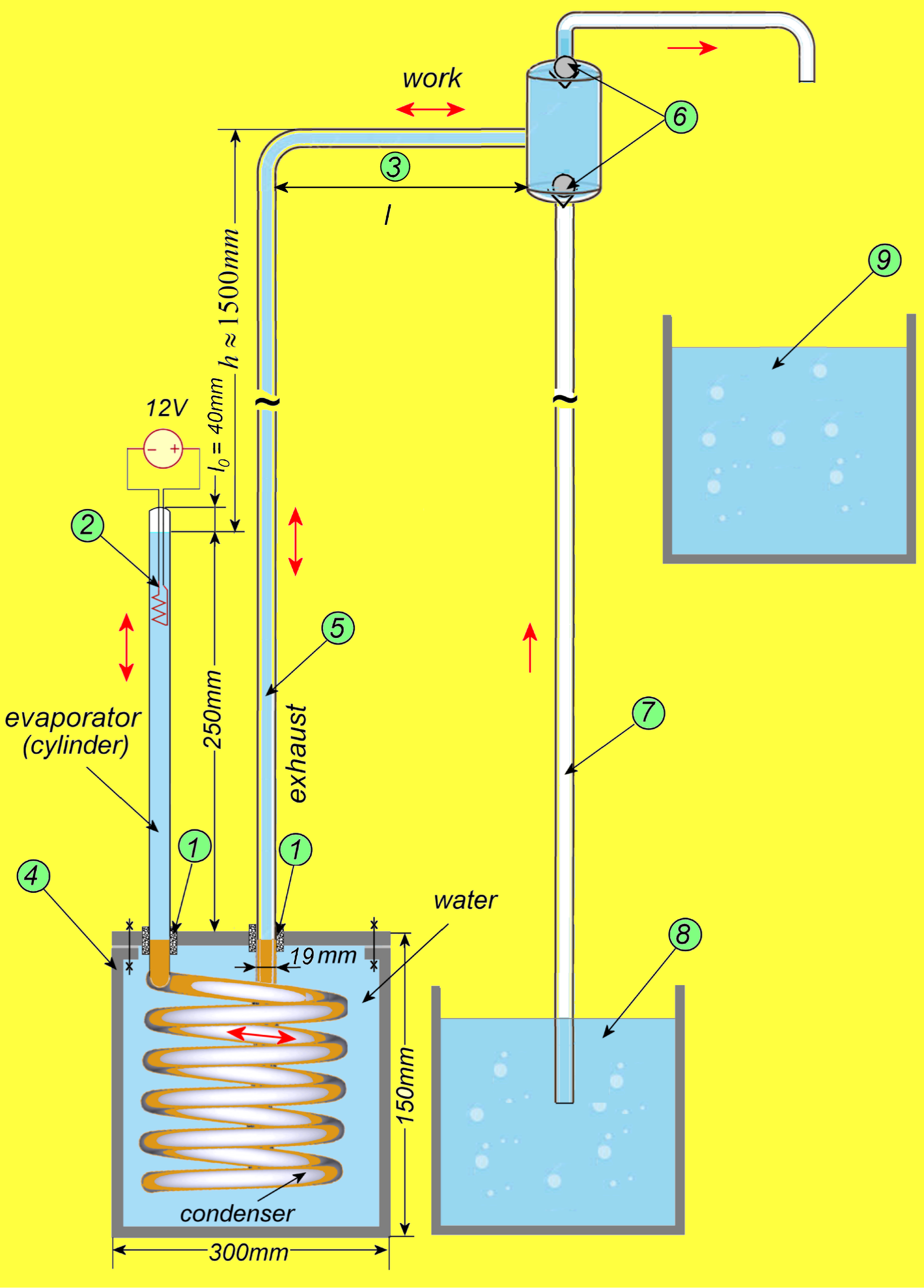}
\caption{- The proposed pump. \numcircledmod{1}: sealing system; \numcircledmod{2}: electrical resistor; \numcircledmod{3}, \numcircledmod{5}: pipes;  \numcircledmod{4} plastic box; \numcircledmod{6}: valves system; \numcircledmod{7}: suction pipe; \numcircledmod{8}: water to be transported; \numcircledmod{9}: transported water.}
\label{fig1}
\end{SCfigure}

The pump was realized experimentally by the author, composed of a cylindrical \emph{evaporator} heated by a resistor, a spiral \emph{condenser} of length $L$, connected with a hose to the \emph{exhaust} (Fig. \ref{fig1}). The system is primed by filling these tubes with water. In the top of the evaporator there remains a small volume of air (vapor) with height $l_0$. The experiment reveals that this vapor volume remains about the same after every admission phase. In the exhaust, the level of the water column, must be higher than that in the evaporator.
Due to the heating and pressure of the water column of height $h$, at some moment, the superheating produces a large amount of superheated saturated vapor which, via the liquid piston, produces work in exhaust. The obtained amount of superheated vapor, once arrived in the condenser, cools and then reduces suddenly its volume\footnote{Due to the short steam cooling time, the vapor condensation is negligible.} and the cycle repeats.

A system of valves  \numcircledmod{6} allows the water suction and evacuation. In this way, the water can be absorbed from the source  \numcircledmod{8} and transported to destination  \numcircledmod{9}. All tubes are \SI{2}{\centi\metre} ($3/4''$) in diameter.
\begin{figure}[!]
\begin{center}
\includegraphics[scale=0.3]{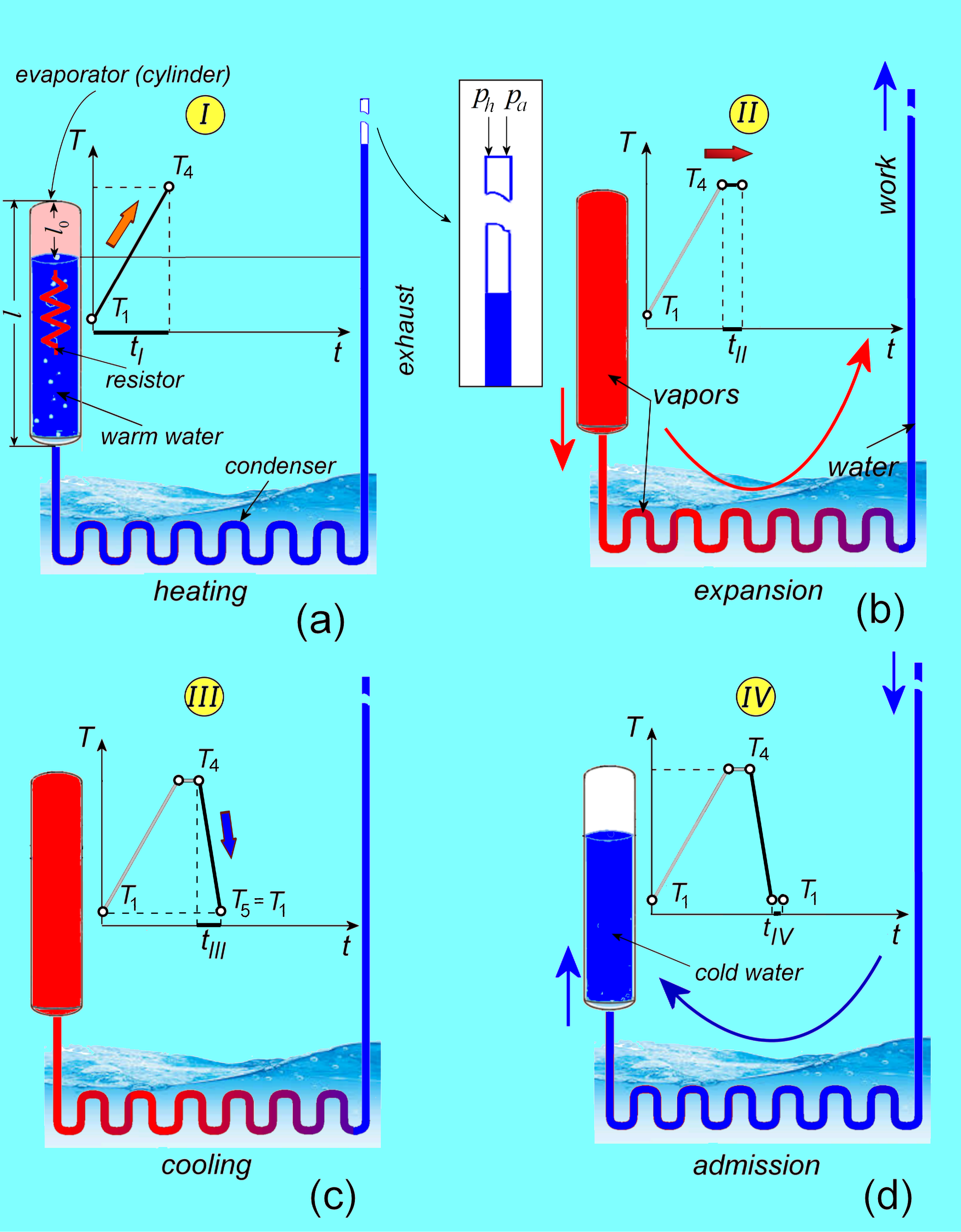}
\caption{Pump phases and cycles time. (a) Heating; (b) Expansion; (c) Cooling; (d) Admission.}
\label{fig2}
\end{center}
\end{figure}
The evaporator, is made from glass in order to observe the phenomena. The resistor  \numcircledmod{2}, submerged in water, is connected to \SI{12}{\volt} AC (or DC) source, which has an electrical resistance of, about \SI{100}-\SI{150}{\ohm}. The condenser is made from copper (copper alloy), to have a quick heat dissipation. The evaporator and the condenser are placed within a sealed plastic box filled with water. The exhaust and suction tubes, \numcircledmod{5} and  \numcircledmod{7} respectively, can be plastic pipes. The tubes are coupled with the sealing system \numcircledmod{1}.

With the pump device dimensions indicated in Fig. \ref{fig1}, the optimum hight $h$ of the exhaust was about \SI{150}{\centi\metre}.

Note that a too high length (water pressure) of the water column in exhaust obstructs the superheating appearance, while a too small value destabilizes the system, in the sense that the volume of water produces a too small pressure to restart the cycle. Also, the length $L$ of the condenser must be long enough to include the entire huge vapor volume obtained by superheating (more than \SI{1}{\metre}). A too small value of $L$ cannot ensure the reducing of the vapor volume created by superheating and, after the superheating phase, the pump stops.

\section{Phenomenological analysis}

\begin{center}
Variables and notations:
\end{center}

\begin{multicols}{2}

Pressure:~ $p[\text{mmHg}]$ ~~or ~~$p[\text{N m}^{-2}]$;

Temperature: ~$T[\si{\celsius}]$ ~~or~~ $T[\si{\kelvin}]$;

Time:~ $t[\si{\minute}]$;

Length:~ $L$, ~~$h$,~~ $r$:  $[\si{\metre}]$ ~~or~~ $[\si{\milli\metre}]$

Volume: ~$V[\si{\metre\cubed}]$ ~~or~~ $V[\si{\litre}]$;

$R=\SI{8.31}{\si{\newton\meter\per\mole\per\kelvin}}$

$c=\SI{4.18}{\kelvin\joule\per\kilo\gram\per\kelvin}$.

Other variables:

Energy: ~$W$, ~~$Q$ measured in~~ $[\si{\joule}]$;

Electrical power: ~$P[\si{\watt}]$;

Electrical resistance: ~$R[\si{\ohm}]$;

Voltage: ~$V[\si{\volt}]$;

Electrical direct source: ~$DC$;

\end{multicols}

\noindent Assume the following:

\noindent -Pressure losses are neglected;

\noindent -The outlet (pushed) water through exhaust and the inlet water through suction tube are equal;

\noindent -Compression and expansion of the working medium are adiabatic reversible (isentropic) processes\footnote{Even the system is not truly isentropic, one considers that the cycle works with ideal gas, where isentropic assumptions are applicable.};

\noindent -Heat losses and pressure drops in particular elements of the system are negligible;

\noindent -After the cooling phase, the water returns into vaporizer, with the same initial temperature $T_1$.

\begin{figure}[b!]
\begin{center}
\includegraphics[scale=0.5]{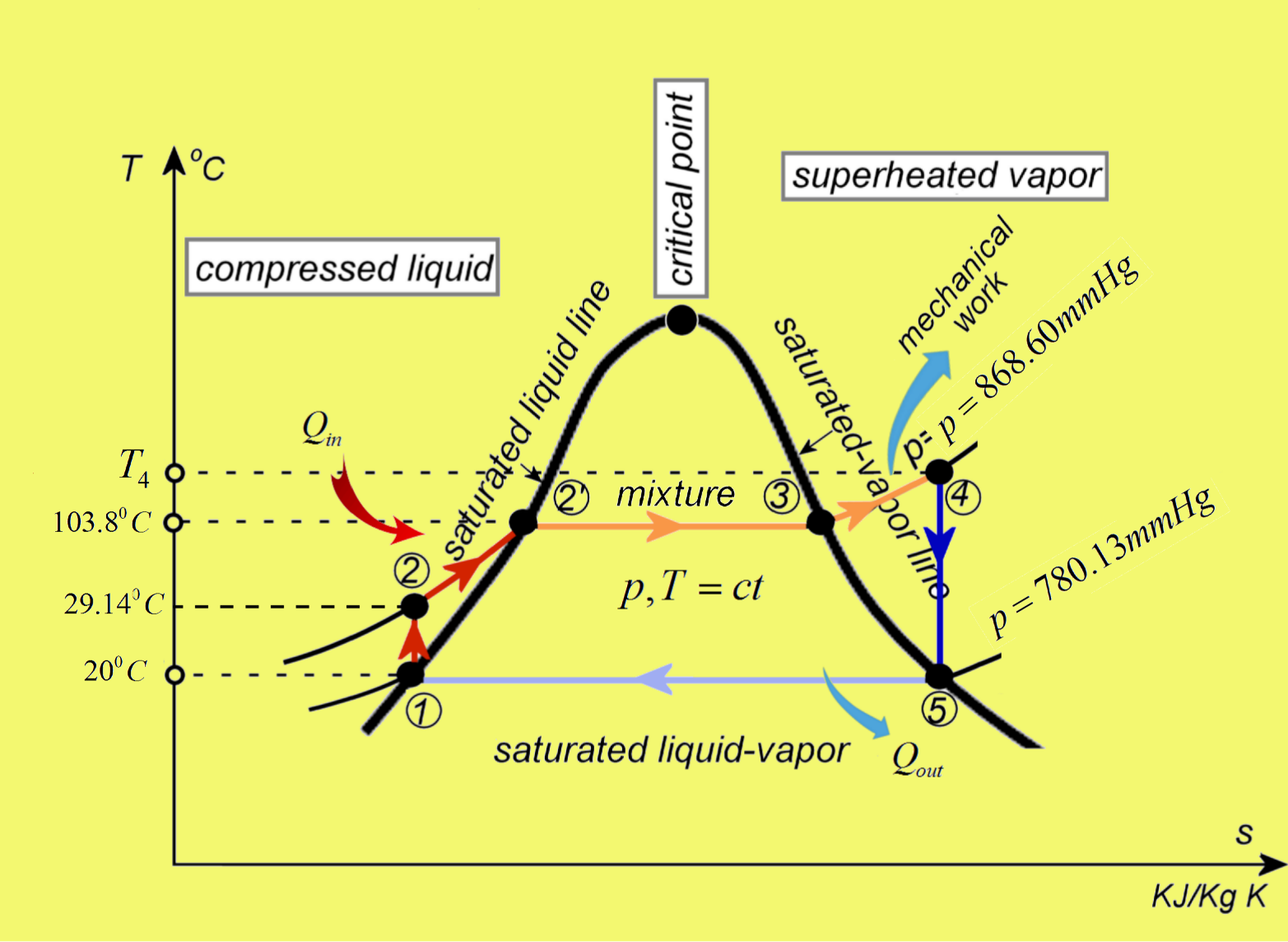}
\caption{$T-s$ diagram of the pump.}
\label{fig3}
\end{center}
\end{figure}

\noindent The proposed pump can be considered as a \emph{four-cycle system} (Fig. \ref{fig2}):

\begin{enumerate}[label=(\arabic*),align=left,leftmargin=0pt,labelindent=\parindent,%
labelwidth=15pt,labelsep=0em,listparindent=\parindent,%
itemindent=\dimexpr\labelindent+\labelwidth+\labelsep-\leftmargin]
\setlength{\itemsep}{0pt}\setlength{\parskip}{0pt}
\makeatletter\@topsep0pt\makeatother 
\item[I:] \textbf{heating} (Fig. \ref{fig2} (a)): for a relative long period of time (the longest, compared to the other times), $t_{\text{I}}$, temperature increases from $T_1$ (point \numcircledmod{1} in Fig. \ref{fig3}) to $T_4$ (superheating temperature at point \numcircledmod{4}; the process of heating of the working medium can be divided into three stages: \emph{heating}, \emph{evaporation} and \emph{overheating};
\item[II:] \textbf{expansion} (Fig. \ref{fig2} (b)): for a short period of time, $t_{\text{II}}$, temperature remains constant, $T_4$, and the work is done;
\item[III:] \textbf{cooling} (Fig. \ref{fig2} (c)): for a short period of time, $t_{\text{III}}$, temperature decreases drastically from superheating $T_{4}$ to $T_1$ and the vapor volume decreases;
\item[IV:] water \textbf{admission} (Fig. \ref{fig2} (d)): for a short period of time, $t_{\text{IV}}$, the water enters into the evaporator and the new water is sucked.
\end{enumerate}

Note that the longest time cycle, $t_I$, corresponds to the heating time (see Fig. \ref{fig2} (a)). This relatively long time (in order of several minutes), represents an disadvantage, which can be compensated by the quantity of transported water, or by the efficiency improvement (see Section 4).

The entire process is reversed by cooling the vapor, and the water from exhaust will go back to vaporizer, retracing the same path. During this process, the amount of heat released
is considered to exactly match the amount of heat added through the heating.

Because of a careful choice of lengths of pump elements (evaporator, condenser, exhaust and connection tube \numcircledmod{3} in Fig. \ref{fig1}), the water circulating within vaporizer-condenser\footnote{Only the exhaust water is (periodically) replaced.} can be considered to be the same; therefore, the pump is a reversible closed system. Actually, the liquid piston can be considered as being composed by the water inside the evaporator and condenser, which pushes the water from the exhaust to the valves subsystem.

The system consists of two isobars and two isentropics processes, and is characterized by the following transformations (see the $\si{T-s}$ diagram in Fig. \ref{fig3}):
    \begin{enumerate}
[align=left,leftmargin=0pt,labelindent=\parindent,%
labelwidth=2pt,labelsep=0em,listparindent=\parindent,%
itemindent=\dimexpr\labelindent+\labelwidth+\labelsep-\leftmargin]
\setlength{\itemsep}{0pt}\setlength{\parskip}{0pt}
\makeatletter\@topsep0pt\makeatother 
\item[\numcircledmod{1}-\numcircledmod{2}]: Reversible isentropic (adiabatic) compression of saturated liquid at the pressure of water column of hight $h$ in the exhaust; a reduced work is produced. The entropy can be reasonably considered as constant (no transfer of heat between the system and the surroundings takes place);
    \item[\numcircledmod{2}-\numcircledmod{2'}-\numcircledmod{3}]: heat addiction: the system receives $Q_{in}$ and an isobaric expansion of steam is produced. At point \numcircledmod{2} the first saturated vapor appears and the system enters the isobar \numcircledmod{2}-\numcircledmod{4}. On this path, the system meets the point \numcircledmod{2'}, where boiling water under the boiling pressure takes place. Next, on the segment \numcircledmod{2'}-\numcircledmod{3}, the system produces a binary mixture liquid-vapor, at constant temperature and pressure. Further heating increment causes evaporation of the liquid until it is fully converted to saturated steam (point \numcircledmod{3}). At the point \numcircledmod{3}, there exists saturated steam under the boiling pressure;
        \item[\numcircledmod{3}-\numcircledmod{4}]: Once the isobar \numcircledmod{2}-\numcircledmod{4} crosses the saturated vapor line (point \numcircledmod{3}), the system enters into the superheating region. Due to further temperature increasing, the isobar ascend to \numcircledmod{4}, where superheating can produce steam by boiling pressure. Further transfer of heat results in an increase in both the temperature and the specific volume;
            \item[\numcircledmod{4}-\numcircledmod{5}]: The highly unstable superheating generates a huge volume of vapor and a work is produced (reversible isentropic expansion);
                \item[\numcircledmod{5}-\numcircledmod{1}]:  constant-pressure and temperature (isobar) heat rejection in the condenser; the system delivers $Q_{out}$. The wet steam condenses completely along the isobar.
\end{enumerate}

For clarity, all important parameters of water and steam, used in this paper, are tabulated in so called ``Steam Tables'' \cite{planet,us,fxsolver,spira,tlv,gestra}.

Suppose that $T_1=\SI{20}{\celsius}$. The pressure at the point \numcircledmod{1} is composed by
the pressure in the top of the evaporator i.e. the atmosphere pressure $p_a$ and the hydrostatic pressure $p_h$ of the (open) exhaust column of the water of height $h$ (Fig. \ref{fig2} (a)):
\begin{equation}\label{pres}
p_1=p_a+p_h=\SI{760}{\mm.HG}+\rho g h =\SI{780.13}{\mm.HG}.
\end{equation}
Since at point \numcircledmod{1} the pressure is given by \eqref{pres}, the necessary temperature to boil the water at the saturated line (point \numcircledmod{2'}) is: $T_2'=\SI{103.8}{\celsius}$ \cite{planet}, and the pressure is: $p_{2'}=\SI{868.61}{mm.Hg}$ \cite{us}. Because the point is on an isobar, the same pressure will be at the point \numcircledmod{2}. Next, one can determine the isentropic \numcircledmod{1}-\numcircledmod{2}, by using the isentropic relation for an ideal gas (see also \cite{fxsolver}):
\begin{equation*}
\frac{T_{2}}{T_1}={\bigg(\frac{p_{2'}}{p_1}\bigg)}^{\frac{\gamma-1}{\gamma}},
\end{equation*}
wherefrom
\begin{equation*}
T_{2}=T_1{\bigg(\frac{p_{2'}}{p_1}\bigg)}^{\frac{\gamma-1}{\gamma}}=\SI{29.13}{\celsius}.
\end{equation*}
From interpolations given by online calculators and Steam Tables (see e.g. \cite{spira},\cite{tlv}, or \cite{gestra}), if one give a temperature slightly above value $T_3$, e.g. $T_4=\SI{104.5}{\celsius}$ (in \cite{spira}, the degrees of superheat for temperature above \SI{103.8}{\celsius} is about \SI{0.2}{\celsius}), one obtains the specific superheated vapor volume $V_4\approx\SI{1.5}{\litre}$\footnote{Along \numcircledmod{2'}-\numcircledmod{3}, the concentration of air in the mixture is about $45\%$ \cite{tlv}.}.

Note that, even the ideal gas law $pV=nRT$ cannot be used for the isentropic transformation, if one considers \SI{1}{\gram} of water participating to this cycle, by this formula, the steam volume is similar to those given by Steam Tables, i.e.,
\begin{equation*}\label{eq}
V_4=\frac{nRT_4}{p_2'}=\frac{{\SI{0.055}{\mole}\times\SI{8.31}{\si{\newton\meter\per\mole\per\kelvin}}\times \SI{377.15}{\kelvin}}}{\SI{115803.80}{\newton\per\meter\squared}}\approx \SI{1.49}{\litre}
\end{equation*}
On the other side, the volume of the exhaust, with $r=\SI{19}{\milli\metre}(3/4'')$ and $h=\SI{1.5}{\metre}$, is $V_{ext}=\pi r^2h\approx\SI{1.7}{\litre}$.

\noindent The necessary mechanical work to pump the \SI{1.5}{\litre} of water at \SI{1.5}{\metre} is
\begin{equation*}\label{cal}
W=Gh=\rho Vgh=\SI{1000}{\kilo\gram\per\metre\tothe{3}}\times\SI{0.0015}{\metre\cubed}\times\SI{9.81}{\metre\per\square\second}\SI{1.5}{\metre}\approx\SI{22.1}{\kilo\gram\metre\squared\per\second\squared}=\SI{22.1}{\joule}.
\end{equation*}
On the other side, to transform one gram of water into steam between $T_1=\SI{20}{\celsius}=\SI{293.15}{\kelvin}$ and $T_4=\SI{104.5}{\celsius}=\SI{377.65}{\kelvin}$, one needs the following energy:
\[
Q_{in}=mc\Delta T=\SI{0.001}{\kilo\gram}\times \SI{4.18}{\kilo\joule\per\kilo\gram\per\kelvin}\times(\SI{377.65}{\kelvin}-\SI{293.15}{\kelvin})\approx\SI{350}{\joule},
\]
\noindent where the specific heat of water is considered as $c=\SI{4.18}{\kelvin\joule\per\kilo\gram\per\kelvin}$.

The underlying electrical power, in about \SI{5}{\minute} necessary for the heating cycle, is $P=W/\SI{5}{\minute}=\SI{1.17}{\watt}$, which can be obtained with \SI{12}{\volt} DC power and electrical resistance of $R=U^2/P=\SI{144}{\volt^2}/\SI{1.17}{\watt}\approx \SI{125}{\ohm}$.

The following efficiency can be determined: $\eta_1=1-W/Q_{in}\approx\SI{94}{\percent}$.
\section{On possible irrigation utilization}
\begin{figure}[!t]
\begin{center}
\includegraphics[scale=0.35]{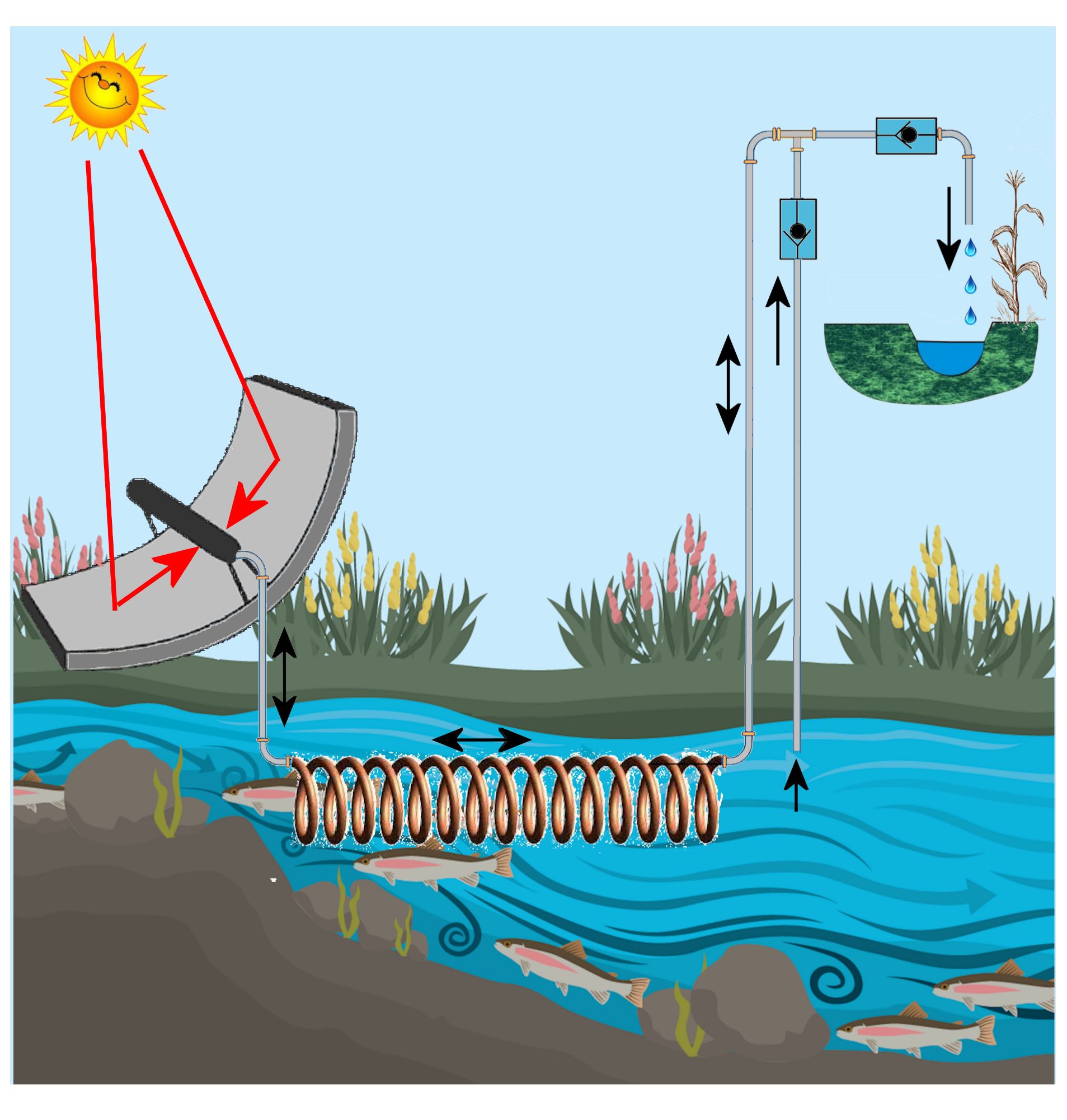}
\caption{Schematic utilization of the proposed pump in irrigation. Solar energy is directed by the concave reflector towards the heating blackened cylinder. The metal (copper) condenser is cooled by the river water. The river water is directed via a pipe system to destination.}
\label{fig4}
\end{center}
\end{figure}

It is easy to understand that, under the sun temperature, larger evaporators can produce superheat and the pump can serve as an irrigation system.

Thus, the evaporator, horizontal and blackened, should be situated in the focus of a concave reflector (concentrator) mirror (Fig. \ref{fig4}), which is a concentrated solar power (see e.g. \cite{11}) and the condenser should be placed within the water source (river). In order to increase the system efficiency, several pumps can be primed into a battery such that, function on evaporator temperature, the cycles of each pump intercalate (start at different time moments) and the entire system work pumps the water continuously-like. A possible intercalation procedure could be as follows: every consecutive pump will start (primed) at the end (or after a fraction) of the previous pump. In this way, because of the periodicity of pump cycles, the delay between two successive expansions (of two nearest pumps) can be reduced substantially and the low efficiency of the pump can be compensated.

\section{Conclusion}
In this paper, a simplified superheating Rankine-like system, without the classical pump and turbine, is presented. Compared to other heating pumps, the exhaust of the proposed pump communicates with the atmosphere pressure.
The experimental model has be heated with an electrical resistance. It was proved both experimentally and analytically that the superheated volume of vapor is large enough to pump water. By the liquid piston of water situated in the exhaust, the vapor produces a mechanical work through pumping water. It is shown that the pump can be utilized as irrigation pump. Moreover, due to its extremely simple structure, the mechanical stresses is drastically reduced.

To increase the efficiency, several pumps can be primed at different instants, so as to work sequentially. Compared to the existing irrigation systems, the advantages of the proposed pump are: cheap with low construction costs (e.g. the evaporator and condenser can be made of metal while the exhaust and pipes can be made of plastic materials), good reliability and reduced maintenance and operating costs.

\raggedright

\end{document}